\documentclass[3p,times,procedia]{elsarticle}
\usepackage{nupha_ecrc}

\volume{00}

\firstpage{1}

\journalname{Nuclear Physics A}

\runauth{Hanlin Li, Jie Zhao, Fuqiang Wang}

\jid{nupha}

\jnltitlelogo{Nuclear Physics A}

\usepackage{amssymb}

\usepackage[figuresright]{rotating}

\usepackage{graphics}

\newcommand {\snn}  {\sqrt{s_{_{\rm NN}}}}
\newcommand {\gevc} {GeV/$c$}
\newcommand {\gevcc}{GeV/$c^2$ }
\newcommand {\pt}   {p_{T}}
\newcommand {\psiRP}    {\psi_{\rm RP}}
\newcommand {\gSS}  {\gamma_{\rm SS}}
\newcommand {\gOS}  {\gamma_{\rm OS}}
\newcommand {\dg}	{\Delta\gamma}
\newcommand {\minv}	{m_{\rm inv}}
\newcommand {\mean}[1]  {\langle #1\rangle}

\begin{document}

\begin{frontmatter}



\dochead{XXVIIth International Conference on Ultrarelativistic Nucleus-Nucleus Collisions\\ (Quark Matter 2018)}

\title{A novel invariant mass method to\\
  isolate resonance backgrounds from the chiral magnetic effect}

\author[addr1]{Hanlin Li}
\author[addr2]{Jie Zhao}
\author[addr2,addr3]{Fuqiang Wang\corref{cor1}}
\address[addr1]{College of Science, Wuhan University of Science and Technology, Wuhan, Hubei 430065, China}
\address[addr2]{Department of Physics and Astronomy, Purdue University, West Lafayette, IN 47907, USA}
\address[addr3]{College of Science, Huzhou University, Huzhou, Zhejiang 313000, China}

\begin{abstract}
  The Chiral Magnetic Effect (CME) refers to charge separation along a strong magnetic field, due to topological charge fluctuations in QCD. Charge correlation ($\dg$) signals consistent with CME have been first observed almost a decade ago. It has also been known since then that the $\dg$ is contaminated by a major background from resonance decays coupled with elliptic flow. 
  In this contribution, we propose differential $\dg$ measurements as function of the pair invariant mass ($\minv$). The $\dg$ in the high $\minv$ region is essentially free of resonance backgrounds. In the low $\minv$ region, the $\dg$ backgrounds show resonance peaks. The CME signal, presumably smooth in $\minv$, may thus be extracted from a two-component model fit. We demonstrate the feasibility and effectiveness of this novel method by using the AMPT and toy-model Monte-Carlo simulations. We also discuss an application of the method in data analysis.
\end{abstract}


\end{frontmatter}


\section{Introduction}
The chiral magnetic effect (CME) is one of the most active research in relativistic heavy-ion collisions~\cite{QM2018}. The CME refers to charge separation along the strong magnetic field produced by the spectator protons~\cite{Kharzeev}. 
Charge separation arises from the chirality imbalance of quarks in local domains caused by topological charge fluctuations in quantum chromodynamics (QCD)~\cite{Kharzeev:1998kz}. Such local domains violate the parity ($\mathcal{P}$) and charge conjugation parity ($\mathcal{CP}$) symmetries~\cite{Kharzeev:1998kz}, which could explain the magnitude of the matter-antimatter asymmetry in the present universe. 

Extensive efforts have been devoted to search for the CME in heavy-ion collisions at RHIC and the LHC~\cite{review,ZhaoReview}. 
The most commonly used observable is the three-point correlator~\cite{Voloshin:2004vk}, $\gamma=\mean{\cos(\alpha+\beta-2\psiRP)}$, where $\alpha$ and $\beta$ are the azimuthal angles of two charged particles and $\psiRP$ is that of the reaction plane. Because of charge-independent backgrounds, the difference $\dg=\gOS-\gSS$ is often used, where $\gOS$ and $\gSS$ refer to opposite-sign (OS) and same-sign (SS) observables, respectively.
There also exist, however, charge-dependent backgrounds, mainly from particle correlations due to resonance decays coupled with the resonance elliptic flow~\cite{Wang:2016iov}: $\gamma\approx\mean{\cos(\alpha+\beta-2\psi_{\rm reso})}\cdot v_{2,{\rm reso}}$. This resonance background was noted by Voloshin~\cite{Voloshin:2004vk} but the quantitative estimate was off by 1-2 orders of magnitude (or a factor of $v_2$)~\cite{Voloshin:2004vk,Wang:2016iov}. When the first experimental data became available~\cite{STAR}, 
it was immediately realized that the data could be largely contaminated by resonance (or cluster) decay backgrounds~\cite{Wang}. 

Particle pair invariant mass ($\minv$) is a common means to identify resonances. In this contribution, we illustrate the invariant mass method~\cite{Zhao:2017nfq,ZhaoReview} and demonstrate that it can be used to measure the CME signal essentially free of resonance backgrounds.

\section{Results}
\begin{figure}[htbp!]
  \centering 
  \includegraphics[width=0.45\textwidth]{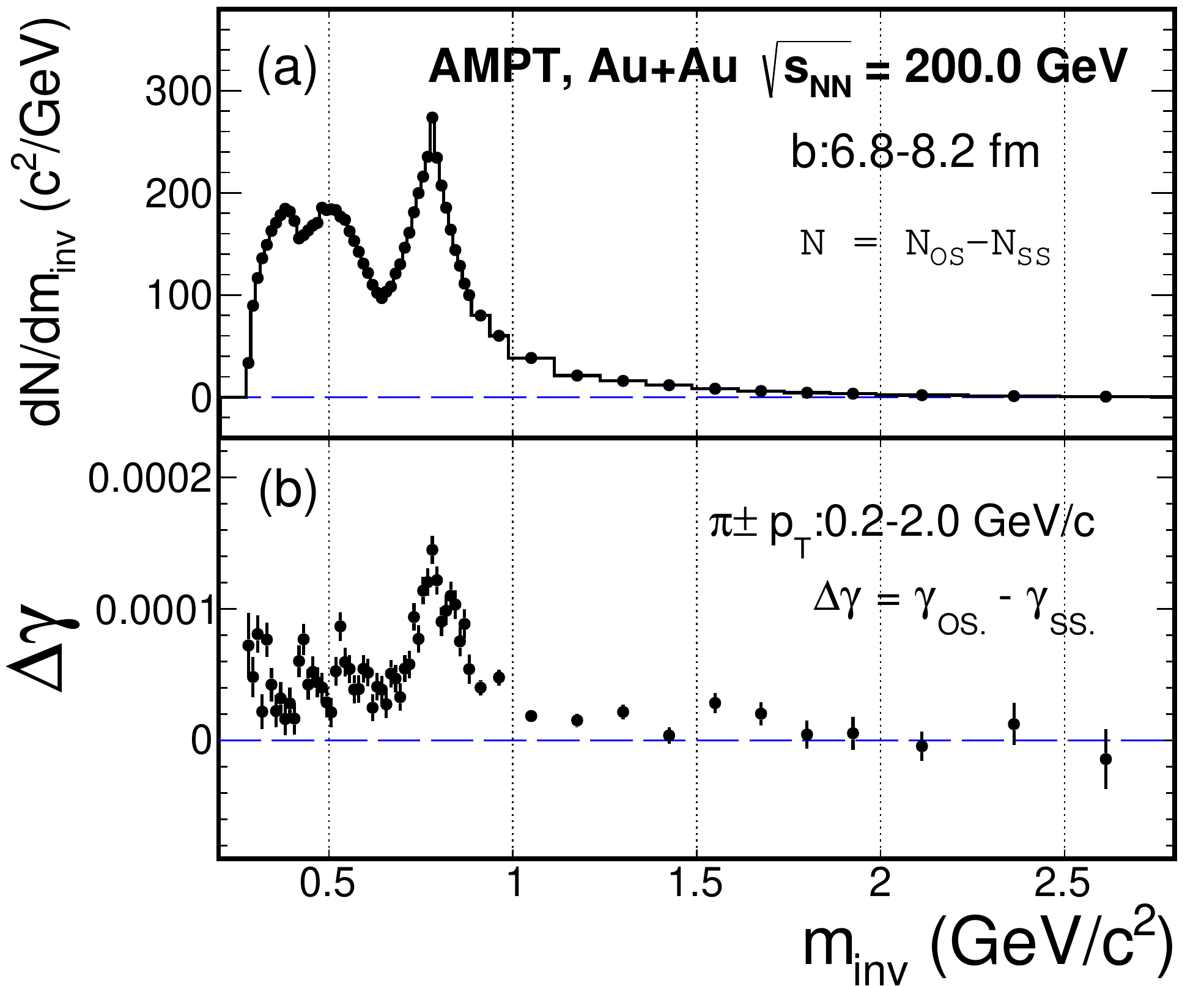}
  \includegraphics[width=0.45\textwidth]{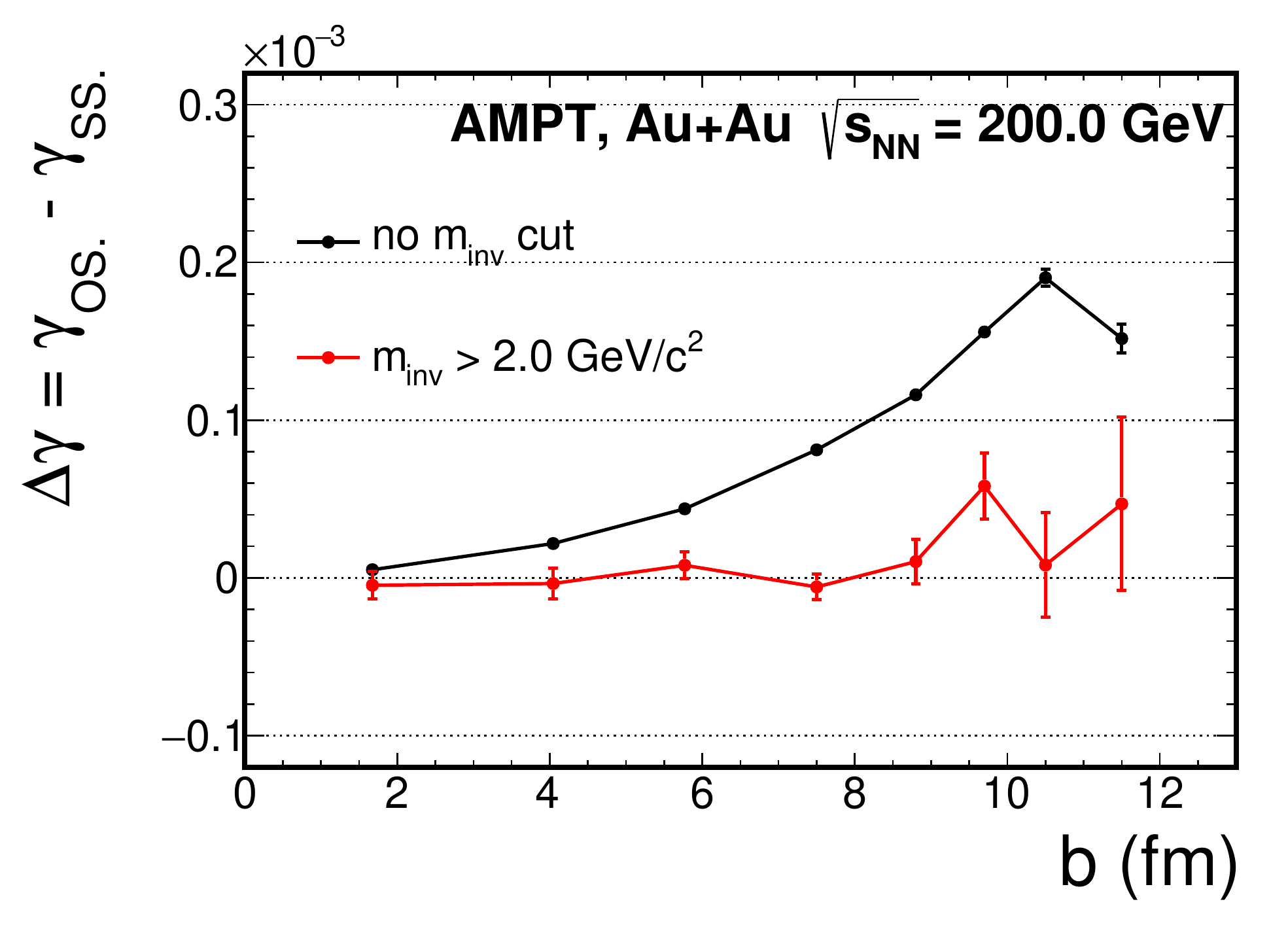}
  \caption{(Color online) AMPT simulation of Au+Au collisions at $\snn=200$~GeV in the impact parameter range of $6.8<b<8.2$~fm. (Upper left) the excess of OS over SS pairs as function of $\minv$; (lower left) $\dg(\minv)$; and (right) the average $\dg$ at $\minv>2$ \gevcc (red) compared to the inclusive $\dg$ (black) as a function of $b$. From Ref.~\cite{Zhao:2017nfq}.}
  \label{fig:ampt}
\end{figure}

We use the AMPT (A Multi-Phase Transport) model to illustrate the invariant mass method. The upper left panel of Fig.~\ref{fig:ampt} shows the execess of OS over SS pairs as a function of $\minv$. The lower left panel shows $\dg(\minv)$. The structures are similar in $r$ and $\dg$; the $\dg$ correlator traces the distribution of resonances. This demonstrates clearly that resonances are the sources of the finite $\dg$ in AMPT.

Most of the $\pi$-$\pi$ resonances are located in the low $\minv$ region~\cite{Agashe:2014kda}. 
It is possible to exclude them entirely by applying a lower $\minv$ cut.
The right panel of Fig.~\ref{fig:ampt} shows the average $\dg$ at $\minv>2$~\gevcc, compared to the inclusive $\dg$ measurement~\cite{Zhao:2017nfq}. The high mass $\dg$ is drastically reduced from the inclusive data. There is no CME in AMPT, and the $\dg$ signal at large mass is indeed consistent with zero. This demonstrates that a lower $\minv$ cut can eliminate essentially all resonance decay backgrounds.

\begin{figure}
  \centering 
  \includegraphics[width=0.72\textwidth]{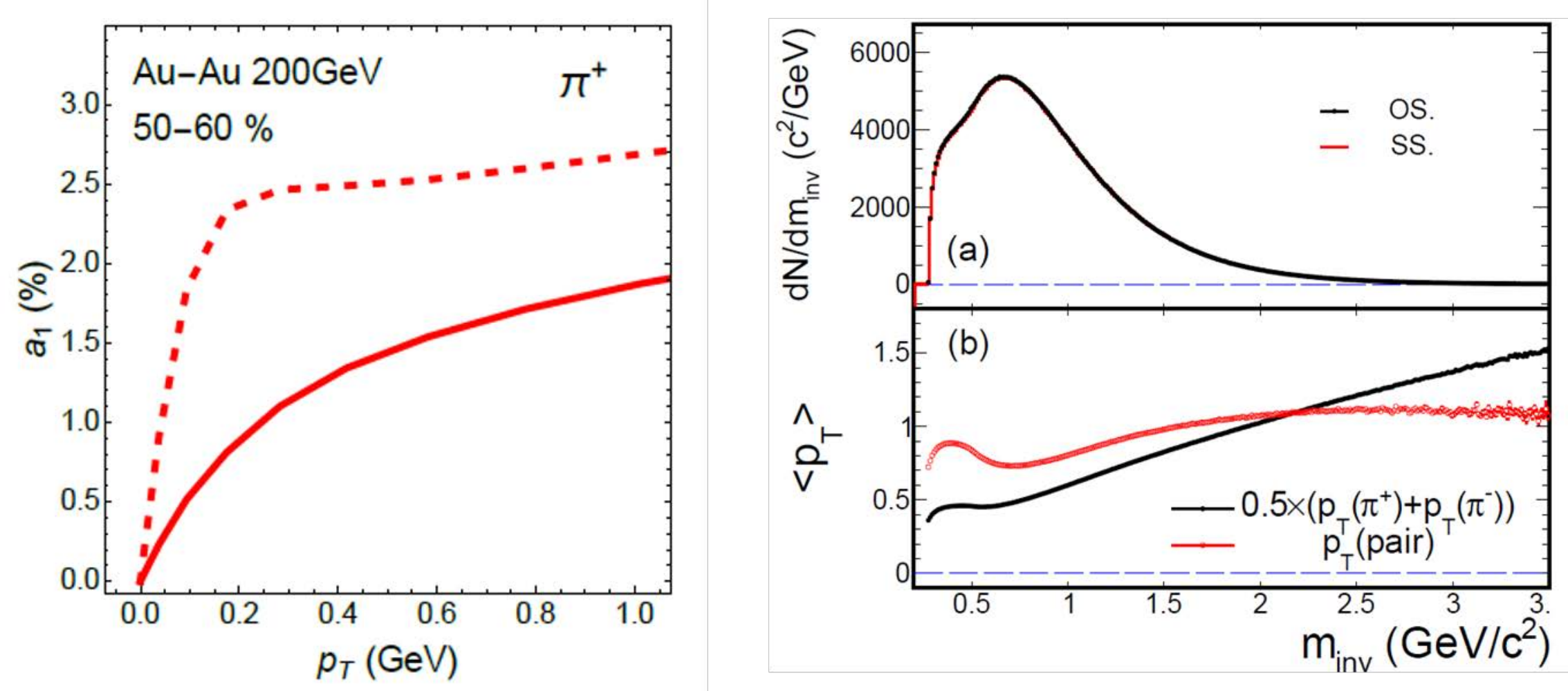}
  \caption{(Color online) Left: the CME charge separation signal strength in directly produced pions (dashed) and in final-state pions (solid) as functions of $\pt$~\cite{Shi:2017cpu}. Upper right: typical $\minv$ distributions of pion pairs in relativistic heavy-ion collisions. Lower right: the $\mean{\pt}$ of single pions (black) and of pion pairs (red) as functions of $\minv$.}
  \label{fig:pt}
\end{figure}

It is generally expected that the CME is a low $\pt$ phenomenon and its contribution to high mass may be small~\cite{Kharzeev,STAR}. 
A recent dynamical model study~\cite{Shi:2017cpu} indicates, however, that the CME signal is rather independent of $\pt$ at $\pt>0.2$ \gevc\ (see Fig.~\ref{fig:pt} left panel), suggesting that the signal may persist to high $\minv$.
The lower right panel of Fig.~\ref{fig:pt} shows the $\mean{\pt}$ of single pions and pion pairs as a function of $\minv$. A cut of $\minv>2$~\gevcc\ corresponds to $\pt\sim 1.2$~\gevc\ which is not very high. The CME signal, if appreciable, should show up in the $\minv>2$~\gevcc\ region.

\begin{figure}
  \centering
  \begin{minipage}{0.5\textwidth}
    \centering 
    \includegraphics[width=0.88\textwidth]{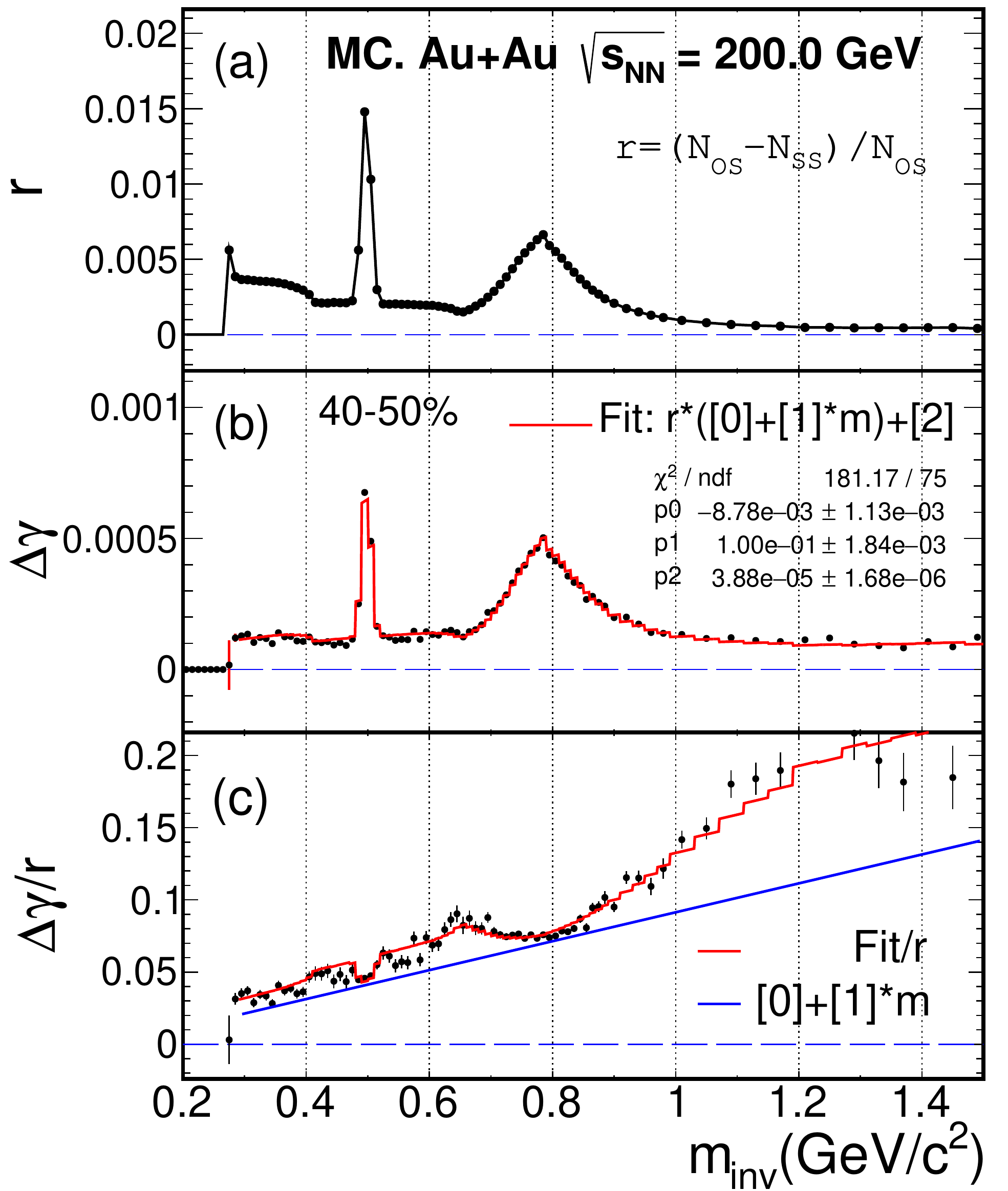}
  \end{minipage}
  \begin{minipage}{0.4\textwidth}
    \caption{(Color online) Toy model simulation from Ref.~\cite{Zhao:2017nfq}. The $\minv$ dependences of (upper) the relative excess of OS to SS pairs, $r=(N_{\rm OS}-N_{\rm SS})/N_{\rm OS}$, (middle) $\dg$, and (lower) $\dg/r$. The red curves are two-component fit. The blue curve in the lower panel is the response function $R(\minv)$ assumed to be linear in the fit.}
    \label{fig:toy}
  \end{minipage}
\end{figure}

One may apply a two-component model~\cite{Zhao:2017nfq},
  $\dg(\minv) \approx r(\minv)R(\minv) + \dg_{\rm CME}(\minv)$,
to extract the possible CME from the low $\minv$ data.
The first term on the r.h.s.~is resonance contributions where the response function $R(\minv)$ is smooth, while $r(\minv)$ contains resonance mass shapes. Consequently, the first term is not ``smooth'' but a peaked function of $\minv$. The second term on the r.h.s.~is the CME signal which should be a smooth function of $\minv$. The $\minv$ dependences of the CME and background are distinct, and this can be exploited to identify CME signals at low $\minv$.
Figure~\ref{fig:toy} shows a toy model simulation including resoances and an input CME signal~\cite{Zhao:2017nfq}. Guided by AMPT input~\cite{Zhao:2017nfq}, the  response function $R(\minv)$ was assumed to be linear. Various forms of $\dg_{\rm CME}(\minv)$ were studied~\cite{Zhao:2017wck}. The two-component model fit is able to extract the input CME signal. 
The lower panel of Fig.~\ref{fig:toy} shows a visual illustration: the ratio of $\dg(\minv)/r(\minv)$ shows a structured modulation on top of a smooth dependence. The structure is due to the ratio of $\dg_{\rm CME}/r$. With the 20\% input CME signal, the inverse structure of $r$ can be visually identified~\cite{Zhao:2017nfq}. 

\begin{figure}
  \begin{minipage}{0.55\textwidth}
    \centering 
    \includegraphics[width=0.78\textwidth]{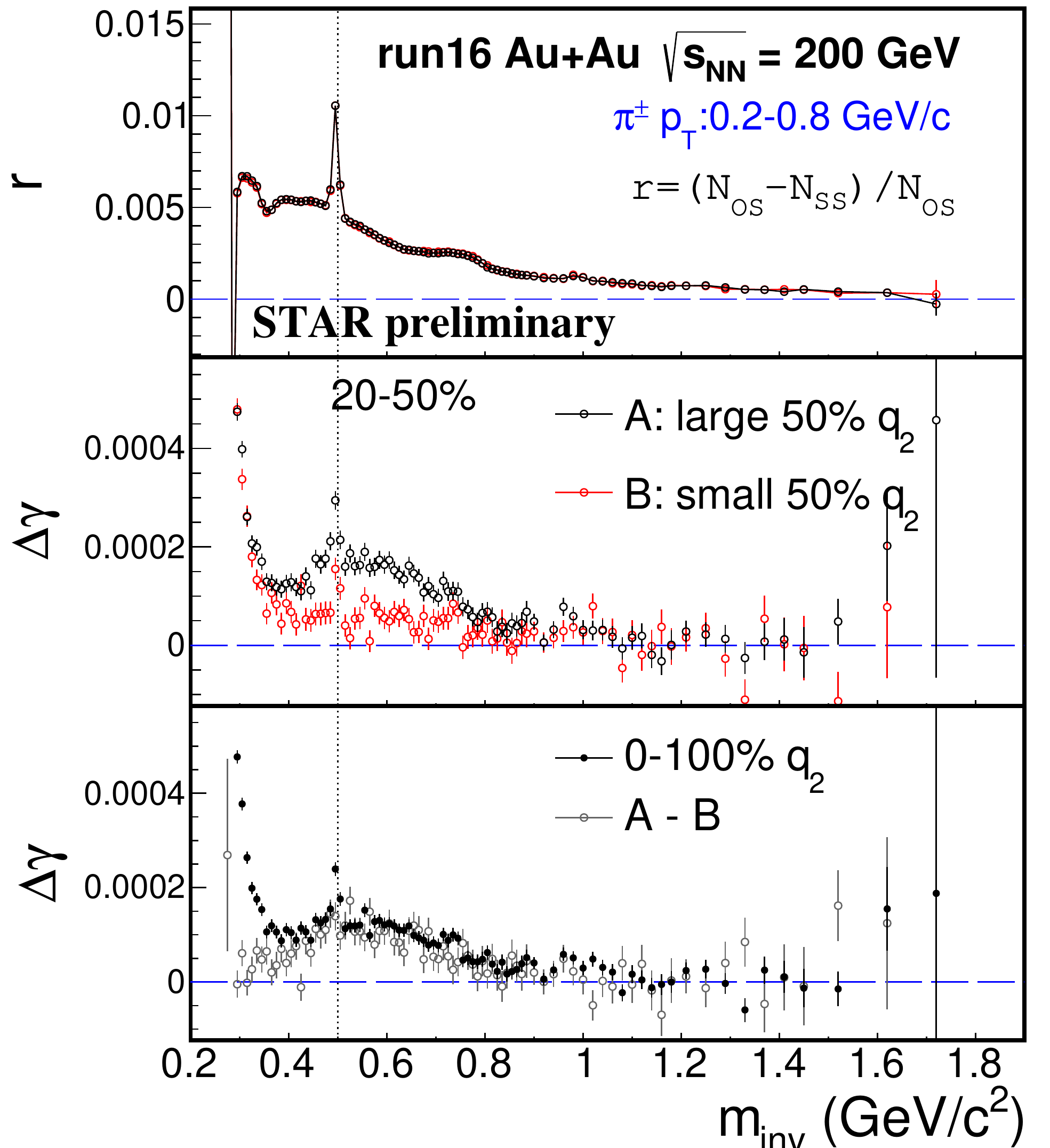}
  \end{minipage}
  \begin{minipage}{0.45\textwidth}
    \includegraphics[width=0.76\textwidth]{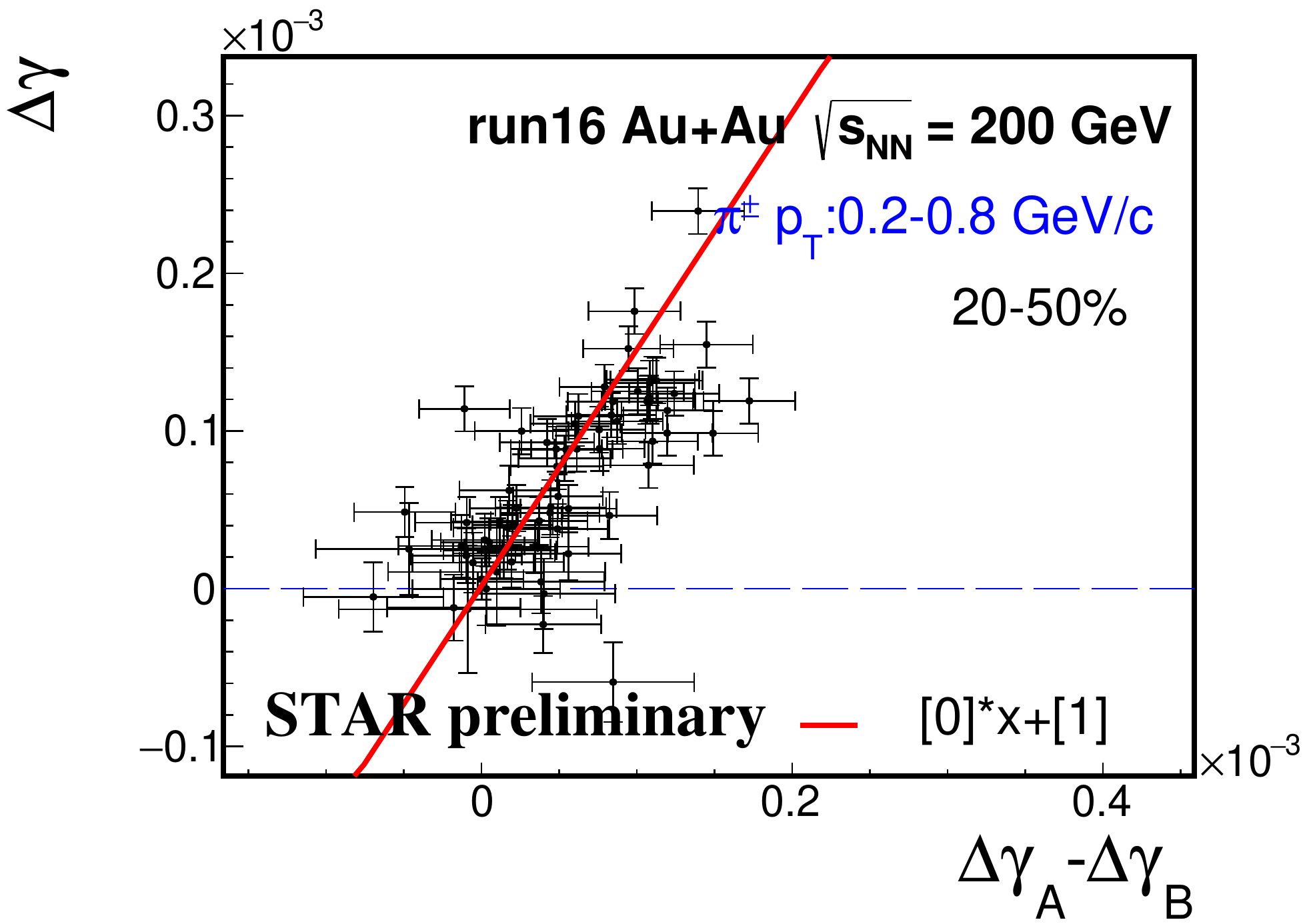}
    \includegraphics[width=0.76\textwidth]{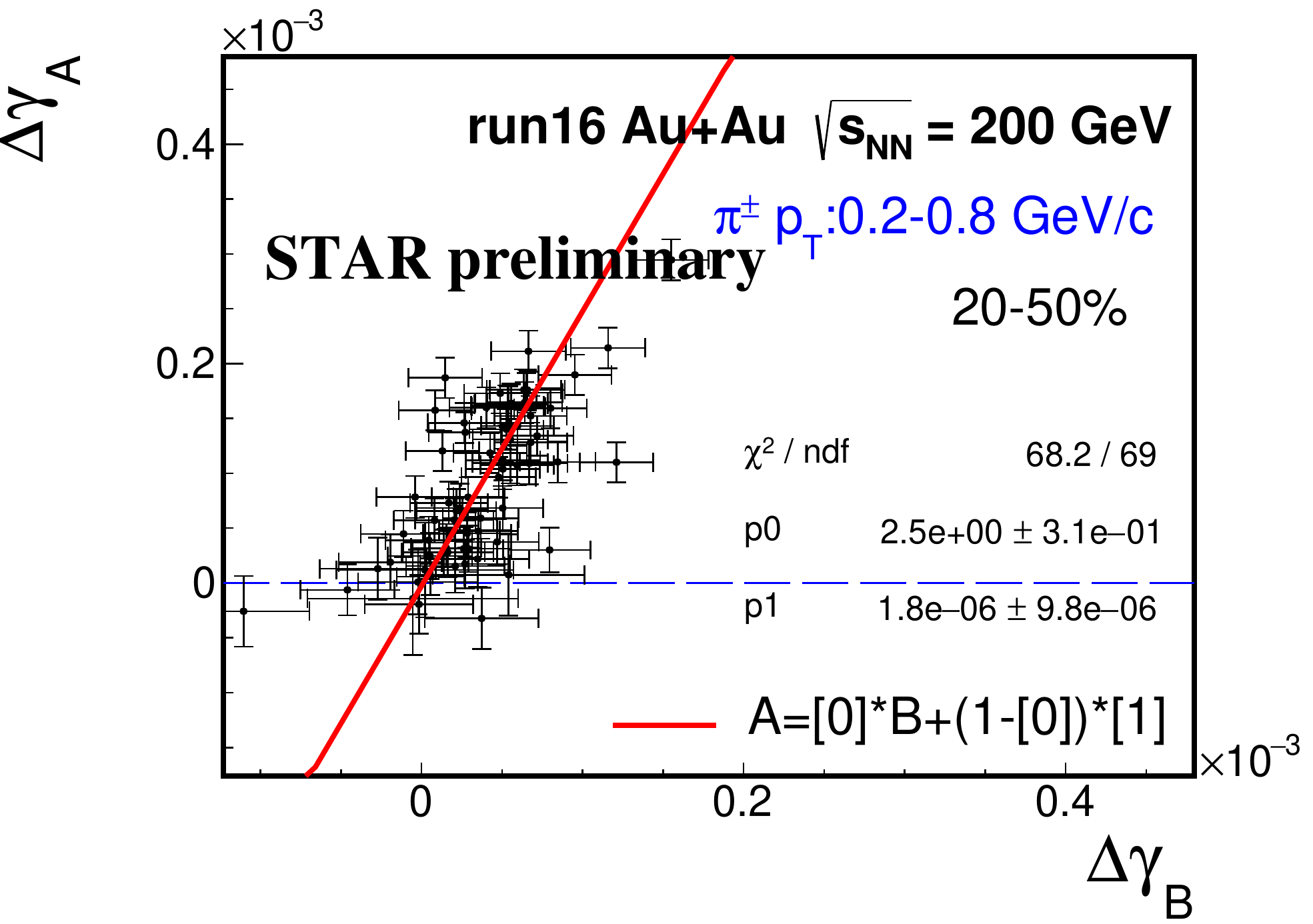}
  \end{minipage}
  \caption{(Color online) Left: the $\minv$ dependences of $r$ (upper), $\dg$ in large and small $q_2$ events (middle), and the $\dg$ difference between large and small $q_2$ events together with the inclusive $\dg$ (lower) in 20-50\% Au+Au collisions at $\snn=200$~GeV from STAR~\cite{ZhaoQM18}. Right: the corresponding $\dg$ versus $\dg_A-\dg_B$ (upper), and $\dg_A$ versus $\dg_B$ (lower). 
    Errors shown are all statistical. From Ref.~\cite{ZhaoQM18}.}
  \label{fig:q2}
\end{figure}

One difficulty above is that the exact functional form of $R(\minv)$ is unknown.
To overcome this difficulty, STAR used the event-shape engineering technique~\cite{ZhaoQM18}, dividing events from each narrow centrality bin into two classes according to the event-by-event $q_2$~\cite{Schukraft:2012ah}. Since the magnetic fields are approximately equal while the backgrounds differ, the $\dg(\minv)$ difference between the two classes is a good measure of the background shape. Figure~\ref{fig:q2} shows $\dg_A$ and $\dg_B$ from such two $q_2$ classes and the difference $\dg_A-\dg_B$ in 20-50\% Au+Au collisions~\cite{ZhaoQM18}. The inclusive $\dg(\minv)$ of all events is also shown. 
With the background shape given by $\dg_A-\dg_B$, the CME can be extracted from a fit $\dg=k(\dg_A-\dg_B)+\dg_{\rm CME}$. Since the same data are used in $\dg$ and $\dg_A-\dg_B$, their statistical errors are somewhat correlated. To propoerly handle statistical errors, one can simply fit the indendent measurements of $\dg_A$ versus $\dg_B$, namely $\dg_A=b\dg_B+(1-b)\dg_{\rm CME}$ where $b$ and $\dg_{\rm CME}$ are the fit parameters. The right panels of Fig.~\ref{fig:q2} show such fits for the STAR Run-16 Au+Au data~\cite{ZhaoQM18}.
Note that in this fit model the background is not required to be strictly proportional to $v_2$. The CME signal is assumed to be independent of $\minv$. The good fit quality seen in Fig.~\ref{fig:q2} indicates that this is a good assumption. 

\section{Summary}
  The Chiral Magnetic Effect (CME) arises from local parity violation caused by topological charge fluctuations in QCD. The CME-induced charge separation measurements by the three-point $\dg$ correlator is contaminated by a major background from resonance decays coupled with elliptic flow. We propose differential $\dg$ measurements as function of the particle pair invariant mass ($\minv$). We show by AMPT and toy-model simulations that (1) $\dg$ in the high $\minv$ region is essentially free of resonance backgrounds, and (2) in the low $\minv$ region, the CME signal may be extracted from a two-component model. We further discuss a data analysis application using the invariant mass method together with event-shape engineering. 

  \vspace*{0.25in}{\bf Acknowledgments.}
  This work was supported in part by U.S.~Department of Energy (Grant No.~de-sc0012910) and National Natural Science Foundation of China (Grant No.~11747312).
  




\bibliographystyle{elsarticle-num}
\bibliography{../../../Papers/ref}







\end{document}